\documentclass[aps,prl,reprint,groupedaddress,showpacs]{revtex4-1}

\usepackage{graphicx}
\usepackage{bm}

\begin{document}


\title{Edge accumulation and currents of moment in  2D topological  insulators}


\author{E.B. Sonin}
\affiliation{Racah Institute of Physics, Hebrew University of
Jerusalem, Jerusalem 91904, Israel}


\date{\today}

\begin{abstract}
The Letter is analyzing bulk spin (moment) currents and spin (moment) accumulation at edges of a 2D topological insulator taking into account reflection from edges. Accumulation occurs only at edge states, which distinguish  topological insulators from conventional ones. The band continuum can support the transverse bulk spin (moment)  current, but it is not governed by the topological Chern number and may exist in a conventional insulator. There is no edge accumulation related with bulk  spin (moment) currents. 
\end{abstract}


\maketitle

\begin{figure}
\begin{center}
   \leavevmode
  \includegraphics[width=0.6\linewidth]{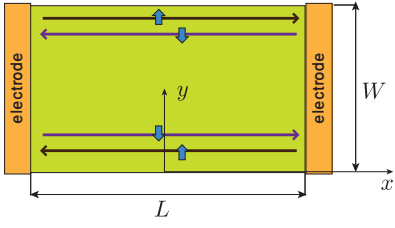} 
 \caption{ (color online) Edge states in a topological insulator. Wide arrows show moment direction (moment quantization axis is not in the plane as in the figure). At the upper edge  rightmovers have moments up while leftmovers have moments down.  At the lower edge directions of motion are opposite. }
 \label{figTI}
 \end{center}
\end{figure}

Topological insulators have recently attracted great attention and their study is developing  into a new exciting area of condensed matter physics. The key signature of 2D topological insulators  is the presence of helical edge states  dictated by topology  \cite{KM,BHZ,Koenig,Joel,Hasan}.  The edge states cross the whole forbidden gap  separating the bulk band continua. Helicity of edge states means that electrons with the same spin can move only in one direction, which is opposite for two  spin directions.  As a result of it,  the edge states are robust against elastic backscattering, which conserves spin, and the electron transport along edges becomes ballistic.  The edge states of the 2D topological insulator were experimentally detected in the HgTe quantum well in studying charge transport  \cite{KoenigE}. It was demonstrated that at the quantum well thickness exceeding the critical value 6.3 nm there was an interval of gate voltages where the conductance reaches the quantum conductance value $2e^2 /h$ independently of the sample width $W$ (Fig.  \ref{figTI}). This is a clear evidence of the ballistic transport along edge states while the main bulk is not conducting. The topological insulators states were also detected in other materials  \cite{Hsi,Xia,Chen}. 

Originally topological insulators were introduced as systems, in which the quantum spin Hall effect (QSH) was expected  \cite{KM,BHZ}, and the state of the topological insulator is frequently called the QSH state. Moreover, sometimes they consider the observed transport properties of ballistic edge states as a manifestation of the QSH effect. In this respect  it is necessary to clarify what could be the exact meaning of the adjective "quantum" added to the spin Hall (SH) effect in 2D topological insulators. Originally the term ``quantum''  was  to stress that the spin conductivity (ratio of the bulk transverse spin current  in the electric field) was  universally determined by the topological Chern number for the 2D Brillouin zone \cite{BHZ}. 
On the other hand, proclaiming that the QSH effect has already been observed, they apparently  refer the word ``quantum''  to the  quantum {\em charge} conductance of the ballistic edge states observed in transport experiments rather than to the quantum bulk {\em spin} conductivity as in the spin-related definition of the QSH effect. Though both, the ballistic edge  states (and the quantum charge conductance as a result of it) and the quantum spin conductivity, originated from topology, the common origin does not make two phenomena identical. A straightforward criterion for a spin-related effect is whether it depends on the spin (or the total moment) value of band electrons. The author is not aware  of any measurement of spin-related properties (spin accumulation and bulk current)   in 2D topological insulators passing this criterion. The theoretical studies of spin properties of 2D topological insulators were restricted with investigations  of the spin conductivity in an infinite sample without any analysis how or whether the value of   the bulk  spin conductivity can be revealed in the experiment. The present Letter focuses on the observability aspect of  spin-related phenomena in a seminfinite topological insulator.

Historically the SH effect was defined as an edge spin accumulation resulting from a bulk spin current normal to an external electric field. But this  definition encountered a problem since spin accumulation is possible without bulk spin current, and, vice versa, spin current not necessarily results in spin accumulation  (see the review \cite{Adv} and references therein). This is a consequence of non-conserved total spin. So the connection of spin accumulation and bulk spin currents is not for granted.  The main outcome of the present analysis is:  (i) spin accumulation in topological insulators  (if and when it were observed) can exist without bulk currents and therefore cannot be a test of  the quantum spin conductivity, and (ii) even direct observation of bulk spin currents by other methods would not mean that they are associated with the Chern number. Bulk spin currents may appear even in the conventional-insulator state with zero Chern number, as the presented analysis of a sample   with a fully reflective border shows.

 In our analysis  we use  the model suggested by Bernevig, Hughes, and Zhang for the topological 
 insulator in the HgTe quantum well \cite{BHZ,Koenig,zhou}. The model is a simplified version of the Kane model and its 4$\times$4 Hamiltonian is given by 
\begin{equation}
{\cal H}=\left(\begin{array}{cc} \hat H({\bm k}) & 0 \\ 0 &  \hat H(-{\bm k})^*\end{array} \right),
   \label{ham4} \end{equation}
where 
\begin{equation}
\hat H({\bm k})=\varepsilon(k)\hat I + d_i \hat \tau _i
       \label{ham2}\end{equation}
is a 2$\times$2 Hamiltonian, $\hat \tau_i$ are Pauli matrices of the pseudospin, $i=x,y,z$,  and $\hat I$ is a unit $2\times2$ matrix. Assuming that all essential processes occur at low $k$ close to the Brillouin zone center, the components $d_i$ are
\begin{equation}
d_x=Ak_x,~~d_y=A k_y,~~d_z=\epsilon_0(k)=M-B k^2.
\end{equation}
Two components of the pseudospin in  any $2\times 2$  block of the Hamiltonian (\ref{ham4}) correspond to the valence (pseudospin up) and the conduction (pseudospin down) bands, which overlap in the  topological insulator phase at $M>0$. The off-diagonal linear in ${\bm k}$ terms in any block  lead to mixing of two original bands and to forming new bands separated by a forbidden gap. The conventional-insulator phase without edge states corresponds to the condition $M<0$.  Further we shall neglect $\varepsilon(k)$ in the Hamiltonian (\ref{ham2}) as not important for the outcome of the analysis \cite{Koenig}.

\begin{figure}
 \includegraphics[width=0.8\linewidth]{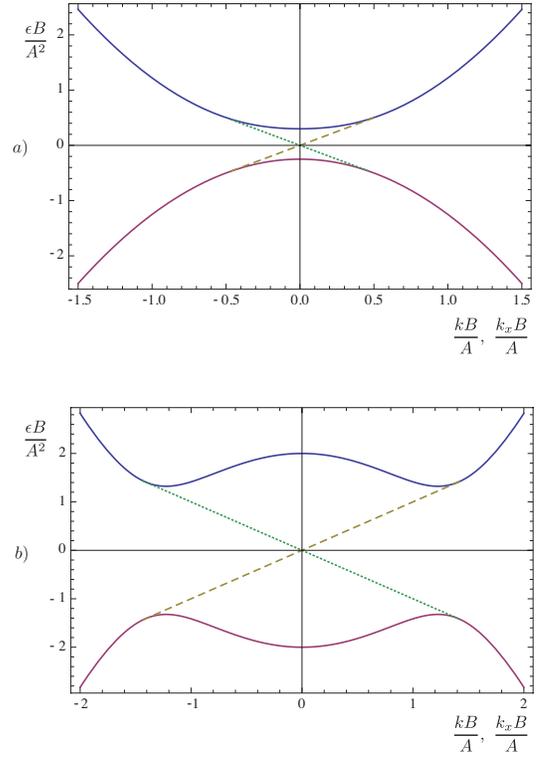}%
 \caption{(color online) Band energy as a function of $k$ (solid lines) and edge state energy as a function of $k_x$ (dotted and dashed lines) in  a topological insulator ($M>0$).  Dashed and dotted lines correspond to opposite moment directions [upper and and lower block of the Hamiltonian (\ref{ham4})].  a) $M=0.3 A^2/B<A^2/2B$; b) $M=2  A^2/B>A^2/2B$.}
\label{f1} \end{figure}

Because of the absence of off-diagonal blocks in  the Hamiltonian (\ref{ham4}), one can analyze states for any block separately. We consider the upper block. The lower block yields the states obtained from those for the upper block by the time-reversal transformation. 
The eigenstates in the ${\bm k}$ space  for the Hamiltonian (\ref{ham2}) are spinors
\begin{eqnarray}
{\bm \Psi}_\pm(k_x,k_y)={1\over \sqrt{2\epsilon}}\left(\begin{array}{c} \sqrt{\epsilon\pm  \epsilon_0}  \\\pm {A(k_x+ik_y) \over \sqrt{\epsilon\pm \epsilon_0}}  \end{array} \right),
\label{spinor} \end{eqnarray}
where $\epsilon=|{\bm d}|=\sqrt{\epsilon_0^2+A^2k^2}$, and the spinors ${\bm \Psi}_+$ and ${\bm \Psi}_-$ correspond to the energies $+\epsilon$ and $-\epsilon$ respectively. 
 At $M<A^2/2B$ the energy of the upper band  has a minimum at $k=0$ (Fig.~\ref{f1}a). At $M>A^2/2B$ the energy has a maximum at $k=0$,  whereas  the  minimum band energy $\epsilon_m = A\sqrt{M/B-A^2/4B^2}$ corresponds to $k_m =\sqrt{M/B-A^2/2 B^2}$  (Fig.~\ref{f1}b).

An edge state near the edge $y=0$ should be a a superposition of  two states of the same energy and  $k_x$: 
\begin{eqnarray}
{\bm \Psi}=\left[a_1{\bm \Psi}_\pm(k_x,k_{y1})e^{ik_{y1}y}+a_2{\bm \Psi}_\pm(k_x,k_{y2})e^{ik_{y2}y}\right]e^{ik_xx},
\nonumber \\
\end{eqnarray}
where $k_{y1}$ and $k_{y2}$ are two complex solutions  with positive imaginary parts of the bi-quadratic equation for $k_y$ following from the energy spectrum at fixed $\epsilon$ and $k_x$.

The condition ${\bm \Psi}=0$ at the edge $y=0$ yields two equations for $a_1$ and $a_2$, which have a solution if
\begin{eqnarray}
\frac{A(k_x+ik_{y1})}{\epsilon_e +\epsilon_0(k_1)}=\frac{A(k_x+ik_{y2})}{\epsilon_e+ \epsilon_0(k_2)}.
\label{DR}\end{eqnarray}
For $M>0$ this  equation is exactly satisfied if $\epsilon_e=-Ak_x$, which corresponds to the edge  state 
\begin{eqnarray}
{\bm \Psi} \propto \left(\begin{array}{c} 1 \\ 1 \end{array} \right) e^{ik_x x}\left(e^{-p_+y}-e^{-p_-y}\right),
 \label{edg}  \end{eqnarray}
where $p_\pm =A/2B \pm \sqrt{k_x^2-M/B+A^2/ 4B^2}$. Considering the lower block of the Hamiltonian (\ref{ham4}) one receives the spectrum $\epsilon_e=Ak_x$.

Starting the analysis of spin currents and accumulation, it is necessary to choose what  ``spin''  we would like to focus on. In HgTe \cite{BHZ,Koenig} the conduction band originates from a  $s$-type ($l=0$) atomic  orbital, and its total moment coincides with spin, whereas the valence band is related to a $p$-type ($l=1$) atomic  orbital and has the total moment $ j=3/2$ with its projection on the quantization axis (the axis $z$ normal to the insulator plane) $m_j=1/2$. This projection of the mechanical total angular momentum but not  genuine spin was in the focus of previous works  \cite{BHZ,Koenig}. However, the mechanical moment would be relevant only if the theory were applied to mechanical effects, like those considered in Ref.~\onlinecite{Mech}. If the goal is to describe electromagnetic phenomena like the Kerr effect, one need the magnetic moment, which depends on the Lande factor of the atomic orbital. So in general the moment projections $s_c$ and $s_v$ on the quantization axis $z$  for the conduction and the valence bands are different, and the operator  of the effective moment is given by 
\begin{eqnarray}
\hat s^z =\bar s^z \hat I +\Delta s^z \hat \tau_z,
   \end{eqnarray}
where $\bar s^z=(s_c+s_v)/2$, and $\Delta s^z = (s_v-s_c)/2$.  If the magnetic moment is studied then $s_c=\mu_B$ and $s_v=2\mu_B/3$, where $\mu_B=e\hbar/2mc$ is the Bohr magneton. 

An interesting consequence of helical edges  in the topological phase $M>0$ is a persistent spin current flowing around the sample. According to Eq.~(\ref{edg}), two  edge states  transport the average spin $\pm \hat s^z$,
and the spin current along the edge is:
\begin{eqnarray}
j^z = \bar s^z(n_\rightarrow+ n_\leftarrow) v_e.
 \label{edCur}  \end{eqnarray}
 where $v_e=d \epsilon_e/\hbar d k_x=A/\hbar$ is the group velocity at edge states. This current exists even in the equilibrium \cite{Butt}, when there is no external electric field and  the 1D densities $n_\rightarrow$ and $n_\leftarrow$ of right-moving and left-moving charge  carriers are equal. So this is one more example of equilibrium spin  currents \cite{Adv}. An electric current $J= e v_e(n_\rightarrow- n_\leftarrow)$ through edge states generated by
 an external electric field leads to moment  accumulation,
\begin{eqnarray}
S_z= {\bar s^z\over e v_e}J,
 \label{Macc}  \end{eqnarray}
without any  bulk moment current. Moreover, bulk moment currents should vanish if 
edge states are robust against elastic scattering. Then they are in the ballistic regime when according to the Landauer--B\"uttiker theory  there is a voltage bias between leads, but no
electric field inside the sample. 

Let us consider now bulk moment currents,  if a finite electric field is present in the bulk (scattering in edge states is possible, or if edge states are absent as in the conventional-insulator state).
The balance equation for the moment (the continuity equation with the torque term) can be derived from the Schr\"odinger equation as explained in details for the Rashba Hamiltonian in Ref.~\onlinecite{Adv}. Restricting ourselves with the $z$ component of the moment density $S_z$, the balance equation is 
\begin{eqnarray}
{\partial S_z\over \partial t} +\nabla_\alpha J^z_\alpha=G^z,
   \end{eqnarray}
where the torque is 
\begin{eqnarray}
G^z=i \Delta s^z A \left\{ {\bm \Psi}^\dagger \cdot  [\vec \nabla\times {\bm \tau}]_z {\bm \Psi}
+  [\vec \nabla\times {\bm \tau}] _z {\bm \Psi}^\dagger \cdot  {\bm \Psi}\right\}
   \end{eqnarray}
and the moment current is given by
\begin{eqnarray}
 J_i^z={1\over 2} {\bm \Psi}^\dagger  \left\{ \hat s^z \hat v_i +\hat v_i \hat s^z \right\} {\bm \Psi} = \bar s^z {\bm \Psi}^\dagger  \hat  v_i {\bm \Psi} + \Delta s^z v_{0i}{\bm \Psi}^\dagger {\bm \Psi}.
\nonumber \\
\label{cur1}   \end{eqnarray}
Here 
\begin{equation}
\hat v_i={\partial \hat H({\bm k})\over \hbar \partial k_i}=v_{0i}(k)\hat \tau_z + A \hat  \tau _i
       \label{cur}\end{equation}
is the  group velocity operator and  $v_{0i}(k)=\partial \epsilon_0/\hbar\partial k_i $. 

The first term in the moment current is proportional to the charge current. Only this term was taken into account in previous publications assuming  $ \Delta s^z=0$. But in general the second term should not be ignored and can be even  crucial. Following the Kubo approach for
calculation of the  moment current one should take into account the electric-field correction to the states replacing   ${\bm \Psi}_\pm(k_x,k_y)$ by the spinors  
\begin{eqnarray}
\tilde{\bm \Psi}_\pm(k_x,k_y)= e^{ik_x x+ik_y y}\left\{{\bm \Psi}_\pm
+{i \hbar eE  \over 4 \epsilon^2}\hat v_x{\bm \Psi}_\pm\right\}.
   \end{eqnarray}
The transverse moment current in this state is  
\begin{eqnarray}
J_y^z(k_x,k_y) = \bar s^z\left(v_y +{i \hbar eE  \over 4 \epsilon^2}{\bm \Psi}_\pm^\dagger[\hat v_y,\hat v_x]{\bm \Psi}_\pm\right)
\nonumber \\
+  \Delta s^z v_{0y}
=\bar s^z\left(v_y +{ eE  \over 4 \hbar}{\cal G}\right)+  \Delta s^z v_{0y},
 \label{trCaur}  \end{eqnarray}
where the term 
\begin{eqnarray}
{\cal G}={A^2\over \epsilon^3}(\epsilon _0-\hbar k v_{0})= \hat{\bm d}\cdot \left[  {\partial \hat{\bm d}\over \partial k_x} \times  {\partial \hat{\bm d}\over \partial k_y}  \right]
      \end{eqnarray}
is responsible for the topological contribution to the current. After integration of the single state current  (\ref{trCaur}) over the $\bm k$ space and  summation of the contributions of the two blocks in the Hamiltonian (\ref{ham4}) with opposite directions of momenta,  only the topological term 
\begin{eqnarray}
\int {\cal G} d{\bm k}=2\pi\left(1+{M\over |M|}\right).
      \end{eqnarray}
contributes to the total moment current. In full agreement with topological theorems, the term appears only in the topological-insulator phase $M>0$ being equal to  the area of the spherical surface subtended 
by the unit vector $ \hat{\bm d}={\bm d}/|{\bm d}|$ over the  2D Brillouin zone (Chern number).

However, this well known result is based on using plane-wave states, ignoring  the boundary conditions on lateral edges. As far as the moment current is proportional to the charge current and there is no moment-flip processes, the reflection boundary conditions totally forbid the transverse current as contradicting the charge conservation. So the bulk current is possible only due to the second term $\propto \Delta s^z$. If the boundary is fully reflective, a proper eigenstate must be a superposition of an incident and a reflected wave: $a_1 \tilde{\bm \Psi}_\pm(k_x,k_y)+a_2 \tilde{\bm \Psi}_\pm(k_x,-k_y)$. In order to satisfy the charge conservation law in the presence of an electric field $|a_1 |^2$ and $|a_2 |^2$ should not be equal.  According to Eq.~(\ref{trCaur}) and  assuming that the electric field does not change  the average density ($|a_1 |^2+|a_2|^2=2$), one obtains that
\begin{eqnarray}
|a_1 |^2= \left(1-{ eE  \over 4 \hbar}{{\cal G}\over v_y}\right),~~|a_2 |^2= \left(1+{ eE  \over 4 \hbar}{{\cal G}\over v_y}\right).
                \end{eqnarray}
As a result, the term in the moment current proportional to the average moment $ \bar s^z$ vanishes but the term proportional to the moment difference $\Delta s^z$ still remains:
\begin{eqnarray}
J_y^z(k_x,k_y) =  \Delta s^z v_{0y}(|a_1 |^2-|a_2|^2)= \Delta s^z { eE  \over 4 \hbar}{\cal G}{v_{0y}\over v_y}.
   \end{eqnarray}
The total current in the whole band does not reduces to the Chern term and is determined by the integral, which does not vanish in a conventional insulator ($M<0$) \cite{com}:
\begin{widetext}
\begin{eqnarray}
\int {\cal G}{v_{0y}\over v_y} d{\bm k}=2\pi \left\{\begin{array}{cc}  \left(-{A\over 2\sqrt{A^2-4MB}}\ln\frac{A^2-2MB+A\sqrt{A^2-4MB} }{A^2-2MB-A\sqrt{A^2-4MB}}
+\ln {A^2-2MB\over 2|M|B}\right) & \mbox{at}~M<{A^2\over 4B} \\  \left(-{A\over \sqrt{4MB-A^2}}\arctan \frac{A\sqrt{4MB-A^2}}{A^2-2M B}
+\ln {A^2-2MB\over 2|M|B}\right) &  \mbox{at}~{A^2\over 4B}<M<{A^2\over 2B}\end{array} \right. .
      \end{eqnarray}
      \end{widetext}
Thus, in contrast to the analysis based on plane-wave eigenstates, the bulk moment current may appear both in the conventional and the topological insulator, and is not governed by the Chern number if the edge of the sample is fully reflective.  However, in the absence of the moment conservation law the bulk current not necessarily leads to accumulation. It may result in an edge torque without accumulation, as in the case of equilibrium spin currents in the Rashba medium \cite{Adv}.  Calculating the accumulated moment with help of eigenstates, which satisfy the boundary conditions, one can see that accumulation takes place only if the distribution among these states has an odd component with respect to $k_x$. This component leads to a longitudinal current. But in a band insulator the odd component is absent since all states in the band are equally filled and there is no longitudinal current. 

In summary, measurement of the moment accumulation at the edge states of the topological insulator if were realized would not provide any information on the bulk moment current. Even direct observation of the bulk moment current would not detect the quantum spin conductivity associated with Chern number in simple geometry with fully reflective edges.  A possible method of bulk moment current detection is observation of an electric field generated by any moving magnetic moment  \cite{Adv,Zhang,Nag}. For example, the edge spin current $j^z$ given by Eq. (\ref{edCur}) leads to the dipole electric field $ \sim (\bar s^z/r^2)(\Delta \epsilon/\hbar c)$, where $r$ is the distance from the edge and $\Delta \epsilon$ is the energy interval, in which edge states are filled.
This is the ``inverse spin Hall effect", which has already been observed but for the diffusion spin current \cite{VT}.  Concluding, the experimental detection of the Chern number, which determines the quantum spin conductivity, seems elusive at the present moment, and some new ingenious set-ups should be looked for this goal.

The author appreciates interesting discussions with A. Bernevig,  B. Laikhtman, J. Moore, G. Refael, and L. Shvartsman.
The work was supported by the grant of the Israel Academy of Sciences and Humanities.

\bibliography{basename of .bib file}

\begin{thebibliography}{99}


 \bibitem{KM} C. L. Kane and E. J. Mele, Phys. Rev. Lett. {\bf 95}, 146802  (2005); {\bf 95}, 226801 (2005).

 \bibitem{BHZ} B. A. Bernevig, T. L. Hughes, and S.-C. Zhang, Science  {\bf 314}, 1757  (2006).
 
 \bibitem{Koenig} M. K\"onig, H. Buhmann, L. W. Molenkamp, T. Hughes, C.-X. Liu, X.-L. Qi, and S.-C. Zhang, J. Phys. Soc. Jpn. {\bf 77},  031007 (2008).
 
 \bibitem{Joel} J. E. Moore, Nature  (London) {\bf 464}, 194 (2010).

\bibitem{Hasan} M. Z. Hasan and C. L. Kane,  arXiv:cond-mat/1002.3895.

  \bibitem{KoenigE} M. K\"onig, S. Wiedmann, C. Br\"une,  A. Roth,  H. Buhmann, L. W. Molenkamp,  X.-L. Qi, and S.-C. Zhang, Science {\bf 318}, 766 (2007).

\bibitem{Hsi} D. Hsieh {\sl et al.}, Science {\bf 323}, 919  (2009).

\bibitem{Xia} Y. Xia {\sl et al.}, Nat. Phys. {\bf 5}, 398  (2009).

\bibitem{Chen}  Y. L. Chen {\sl et al.}, Science {\bf 325}, 178  (2009).

\bibitem{Adv} E. B. Sonin, Adv. Phys. {\bf 59}, 181  (2010). 

\bibitem{zhou} B. Zhou, H.-Z. Lu, R.-L. Chu, S.-Q. Shen, and Q. Niu, Phys. Rev. Lett. {\bf 101}, 246807 (2008).

\bibitem{Mech} E. B. Sonin,  Phys. Rev. Lett.  {\bf 99}, 266602 (2007).

\bibitem{Butt} M. B\"uttiker, Science {\bf 325}, 278  (2009).

\bibitem{com} One cannot extend this analysis on $M>A^2/2B$ since in this case there are two real values of $k_y^2$  for the same energy (see Fig.~\ref{figTI}b). So there are two reflected waves,  and the moment current cannot be calculated without a detailed analysis of the wave function near the edge. 

  \bibitem{Zhang} S. Zhang, J. Appl. Phys. {\bf 89}, 7564 (2001).

  \bibitem{Nag}  N. Nagaosa, J. Phys. Soc. Jpn. {\bf 77},  031010 (2008).
  
  \bibitem{VT} S. O. Valenzuela and M. Tinkham, Nature (London) {\bf 442}, 176 (2006).


\end{thebibliography}

\end{document}